\documentclass[12pt]{article}
\usepackage{epsfig}
\usepackage{cite}
\newlength{\dinwidth}
\newlength{\dinmargin}
\begin{document}
\newcommand{\be}{\begin{equation}}
\newcommand{\ee}{\end{equation}}
\newcommand{\ba}{\begin{eqnarray}}
\newcommand{\ea}{\end{eqnarray}}
\newcommand{\tdm}[1]{\mbox{\boldmath $#1$}}
\def\qqd{(q\bar q)_{\rm\small dipole}}
\newcommand{\bbbar}{$b\bar{b}$ }
\newcommand{\ccbar}{$c\bar{c}$ }

\titlepage
\begin{flushright}
{\sc TSL/ISV-2002-0265     \\
November 12, 2002}       \\
\end{flushright}
\begin{center}
\vspace*{2cm}
{\Large\bf Unintegrated gluon in the photon \\
and heavy quark production}

\vspace*{1cm}
 L. Motyka$^{a,b}$ and N. T\^\i mneanu$^{a}$ 
\vspace*{0.5cm}
\begin{center}
{$^{a}$ High Energy Physics, Uppsala University,
Box 535, S-751~21 Uppsala, Sweden}   \\
{$^{b}$ Institute of Physics, Jagellonian University,
Reymonta 4, 30-059 Krak\'{o}w, Poland}\\

\end{center}
\end{center}
\vspace*{1.5cm}
\vspace*{2cm}
\begin{abstract}

The unintegrated gluon density in the photon is determined, 
using the Kimber-Martin-Ryskin prescription . 
In addition, a model of the unintegrated gluon is 
proposed, based on the saturation model extended to the 
large-$x$ region. These gluon densities are applied to obtain
cross sections for charm and bottom production in $\gamma^* p$ and 
$\gamma\gamma$ collisions using the $k_t$ factorization 
approach. We investigate both direct and resolved photon
contributions and make comparison with
the results from the collinear approach and the experimental data.
An enhancement of the cross section due to inclusion
of non-zero transverse momenta of the gluons is found. 
The charm production cross section is consistent with the 
data. The data exceed our conservative estimate for bottom 
production in $\gamma p$ collisions,  but theoretical uncertainties
are too large to claim a significant inconsistency. 
A substantial discrepancy between theory and the experiment
is found for $\gamma\gamma \to b\bar b X$, not being cured by
the $k_t$ factorization approach. 
\end{abstract}
\newpage

\section{Introduction}

Production of heavy quarks at high energies has 
been vigorously studied experimentally in recent
years. Measurements of cross sections for charm 
and bottom production have been performed in
proton-proton \cite{exp_pp}, proton-photon 
\cite{exp_gp,exp_f2c,EMCBB,HERABB,ZEUSBB} 
and photon-photon \cite{LEPCC,LEPHQ} collisions.  
The charm data in $\gamma^* p$ and $\gamma\gamma$ 
collisions may be reasonably well described
by the standard collinear formalism, based on LO~QCD with 
NLO corrections \cite{charm_th}. For bottom however,
the experimental results exceed significantly the
theoretical expectations in all cases. The largest 
discrepancy has been found for bottom production 
in $\gamma\gamma$ collisions at LEP \cite{LEPHQ}, where 
the measured cross section is larger by a factor of 
four than the QCD prediction. 

The enhancement of the bottom production cross section 
was reported with different beams and at different
energies, which suggests the presence of an important systematic 
effect,  omitted in the QCD analysis. It is particularly
puzzling because of the large mass of the bottom quark,
giving a safe ground for the perturbative approach.  
Attempts have been made \cite{CE,CCH,LRSS,RSS,LSZ,LSZ2,HJ,Szczurek,Cristiano} 
to resolve this problem
by going beyond the standard collinear formalism 
and use the $k_t$ factorization approach \cite{CE,CCH,LRSS,kt_fac}.
Thus, instead of assuming that massless partons are 
distributed in the colliding objects having a negligibly 
small transverse momentum, one considers
the complete kinematics of parton scattering,
taking into account the transverse momenta.

This intrinsic transverse momentum $\tdm k$ of the parton is built
up in the perturbative evolution, as a result of
subsequent emissions of gluons or quarks and its
distribution is parameterized by the unintegrated 
parton distribution. The influence of the parton transverse 
momentum on the cross section depends on the relevant 
hard matrix element, which has to be evaluated for virtual
partons (off-shell matrix element). 
Calculations using the off-shell matrix
elements combined with the unintegrated parton
distributions were performed for bottom production
in $p\bar p$ collisions\cite{HJ,RSS,LSZ2}. Indeed, the obtained 
cross sections are larger than those in the collinear approximation
and agree with the data within uncertainties. For $\gamma^* p$ an
enhancement is also found for a direct photon\cite{HJ,LSZ,Cristiano}, 
but is not sufficient to describe the data.

The unintegrated gluon distribution in the proton evaluated 
at the factorization scale $\mu$ ${\cal F}_g(x,\tdm k ^2,\mu^2)$ 
is a subject of intensive studies itself (for a review, see \cite{smallx}).
This quantity depends on more degrees of freedom  
than the collinear parton density, and is therefore less constrained 
by the experimental data. Various approaches to model the 
unintegrated gluon have been proposed. 
For instance, in the leading logarithmic~$1/x$ approximation, 
evolution of ${\cal F}_g(x,\tdm k^2,\mu^2)$ is given by the 
BFKL\cite{BFKL} or CCFM\cite{CCFM} equations. 
The unintegrated gluons following from those equations were 
fitted successfully to inclusive data from $ep$ scattering 
\cite{KMST,JS}.
This approach is restricted to the small~$x$ regime.
Recently, it has been shown \cite{KKMS,KMR,KMR-pre} that the 
information contained in the collinear parton densities
combined with the properties of parton
emission amplitudes (e.g.\ the angular ordering) is sufficient 
to determine the unintegrated gluon up to a large~$x$.  
An interesting model for the gluon is also given
by the successful saturation model\cite{GBW},
introduced by Golec-Biernat and W\"usthoff (GBW).

Models for the unintegrated parton distributions in the photon
were not available until very recently\cite{Lund,SFDP,CCFM-photon} and 
no results for the resolved photon are known beyond the collinear 
limit. Thus, the main purpose of this study is to obtain
in an independent way, the unintegrated gluon distributions 
in the photon, using the Kimber-Martin-Ryskin (KMR) \cite{KMR} prescription 
and to apply them to describe the heavy quark
production (charm and bottom) in  $\gamma\gamma$ 
and $\gamma p$ collisions.
Application of the $k_t$ factorization formalism
for the case of resolved photon(s) is performed for 
the first time.
We also obtained an alternative gluon density in the photon
based on the generalized saturation model \cite{TKM}
for $\gamma\gamma$ interactions. 
We will explore a variety of gluon parameterizations
and account the other model ambiguities in order
to estimate the theoretical uncertainties.
We will examine whether the excess of the bottom production 
in these processes can be explained within the $k_t$ factorization 
approach.

The paper is organized as follows: in Section~\ref{sec2} the unintegrated
gluon in the photon is obtained and its properties
are discussed, in Section~\ref{sec3} the $k_t$ factorization formulae
are presented and in Section~\ref{sec4} the cross sections for 
heavy quark production in $\gamma\gamma$ and 
$\gamma p$ collisions are calculated.
A discussion of the results is given in Section~\ref{sec5},
followed by conclusions in Section~\ref{sec6}.

\section{Unintegrated gluon distributions in the photon}
\label{sec2}

In the construction of unintegrated gluon distributions in the photon we apply 
the same method as for the distributions in the proton. The off-shell parton 
distributions in the proton are better known and their properties have been 
investigated to a great extent (see \cite{smallx} and references there-in),
while similar distributions in the photon are poorly known and no attempts have
been made to describe them until recently \cite{Lund,SFDP,CCFM-photon}.

\subsection{The KMR approach}
\label{sec2a}

The conventional gluon distribution $g(x,\mu^2)$ corresponds to the density of
gluons in the photon having a longitudinal momentum fraction $x$ at the
factorization scale $\mu$. This distribution  satisfies the
Dokshitzer-Gribov-Lipatov-Altarelli-Parisi (DGLAP) evolution 
\cite{DGLAP} in $\mu^2$ and it is universal for the photon in different 
processes.
The distribution does not contain information about the  transverse momenta 
$\tdm k$ of the gluon, which is integrated over up to the
factorization scale $\mu
$\be
x g(x,\mu^2) = \int^{\mu^2} d\tdm k^2 {\cal F}_g(x,\tdm k^2,\mu^2).
\label{xgx}
\ee
However, in order to better describe processes by properly considering
the transverse momentum of the gluon, unintegrated parton
distributions ${\cal F}_g(x,\tdm k^2,\mu^2)$ are needed. These distributions
take into account the complete kinematics of the partons entering the process
at the leading order (LO).

As a first attempt,
the unintegrated gluon density may simply be obtained \cite{KKMS}, at very low 
$x$, from the collinear gluon density $x g(x,\mu^2)$. This density becomes 
independent of the hard scale $\mu^2$, and will only  depend on one scale
$\tdm k^2
$\be
{\cal F}_g(x,\tdm k^2,\mu^2) = 
\left. \frac{\partial}{\partial Q^2} [x g(x,Q^2)] \right|_{Q^2=\tdm k^2} .
\label{deriv}
\ee

The above equation no longer holds, as $x$ increases, since 
${\cal F}_g(x,\tdm k^2,Q^2)$ becomes negative.  
This may be circumvented, however, by introducing a 
Sudakov form factor $T_g(Q,\mu)$, which takes into account subleading
corrections at low $x$. Thus, the unintegrated distribution has now a 2-scale 
dependence \cite{DDT,KMR-pre}

\be
{\cal F}_g(x,\tdm k^2,\mu^2) =\left.  \frac{\partial}{\partial Q^2} [x g(x,Q^2) 
\times T_g(Q,\mu)] \right|_{Q^2=\tdm k^2} ,
\label{derivsudakov}
\ee
with  the form of $T_g(Q,\mu)$ given below. The form factor represents the 
probability of the gluon with the transverse momentum $\tdm k$ to survive
untouched in the evolution up to the factorization scale.

A better framework for unifying the small $x$ and large $x$ regions is provided
by the Catani-Ciafaloni-Fiorani-Marchesini (CCFM) equation \cite{CCFM}. This
equation is an evolution equation for the unintegrated gluon distribution
${\cal F}_g(x,\tdm k^2,\mu^2)$ which considers real gluon 
emission in a ladder and is 
based on angular ordering of the gluons in the chain. The formalism has a 
natural interplay of two scales: the transverse momentum $\tdm k$ of the gluon and 
the hard scale $\mu$, which corresponds to the maximal angle of emitted gluons. 
Thus, the unintegrated
gluon distributions which can be constructed will have a 2-scale dependence
${\cal F}_g(x,\tdm k^2,\mu^2)$, where $\mu$ will have a dual role, that of factorization
scale and controlling the angular ordering. At small $x$, the CCFM formalism
is equivalent, in the leading $\log(1/x)$ approximation, to the 
Balitskij-Fadin-Kuraev-Lipatov (BFKL) formalism\cite{BFKL}, and
${\cal F}_g(x,\tdm k^2,\mu^2)$ which satisfies the BFKL equation becomes
$\mu$-independent. At moderate $x$, the angular ordering is replaced by $k_t
$ordering, and the CCFM equation reduces to DGLAP.

A simplified solution to the complicated 2-scale CCFM evolution was obtained
by Kimber, Martin and Ryskin (KMR) in \cite{KMR}. They observed that the $\mu
$dependence in the distributions enters only in the last step of the evolution,
and single-scale evolution equations can be used up to the last step.
In this approximation, the unintegrated gluon distribution is given by
\be
{\cal F}_g(x,\tdm k^2,\mu^2) = \frac{T_g(\tdm k,\mu)}{\tdm k^2} \frac{\alpha_s(\tdm k^2)}{2\pi}
\int_x^{1-\delta} dz \left[ P_{gg}(z) \frac{x}{z}g(\frac{x}{z},\tdm k^2) 
+ \sum_q P_{gq}(z) \frac{x}{z}q(\frac{x}{z},\tdm k^2)\right] ,
\label{kmreq}
\ee
where $P_{gg}(z)$ and $P_{gq}(z)$ are the gluon and the quark 
splitting functions, while $g(x,Q^2)$ and $q(x,Q^2)$ are the 
conventional gluon and quark densities. 
The Sudakov form factor introduces the dependence on the second scale $\mu
$in the last step of the evolution and has the following form

\be
T_g (\tdm k,\mu) = \exp\left( -\int_{\tdm k^2}^{\mu^2} \frac{dp^2}{p^2} 
\frac{\alpha_s(p^2)}{2\pi} 
\int_0^{1-\delta} dz z \left[P_{gg}(z)+\sum_q P_{qg}(z)\right]\right) ,
\label{sudakov}
\ee
where $\delta = \frac{p}{p+\mu}$ is chosen to provide the correct angular
ordering  of the real gluon emissions.

\begin{figure}[th]
\begin{center}
\begin{tabular}{cc}
\epsfig{width= 0.45\columnwidth,file=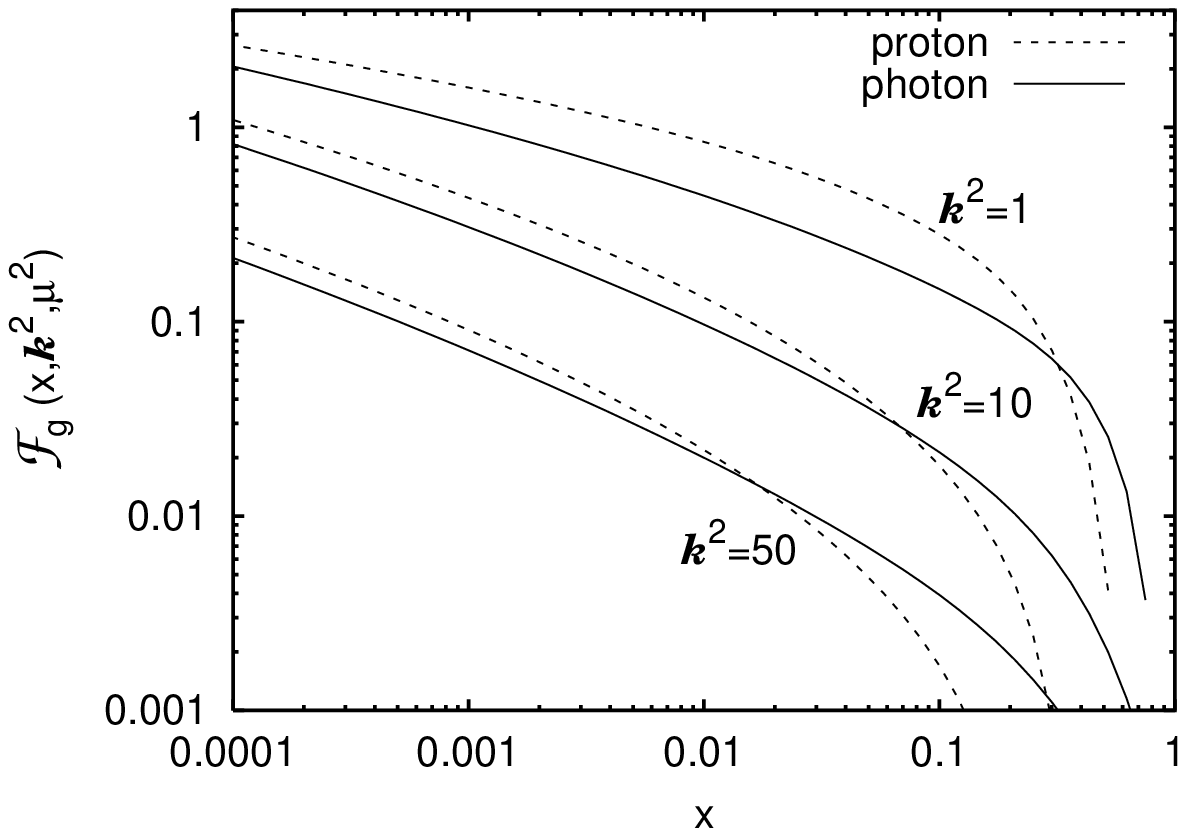} &  
\epsfig{width= 0.45\columnwidth,file=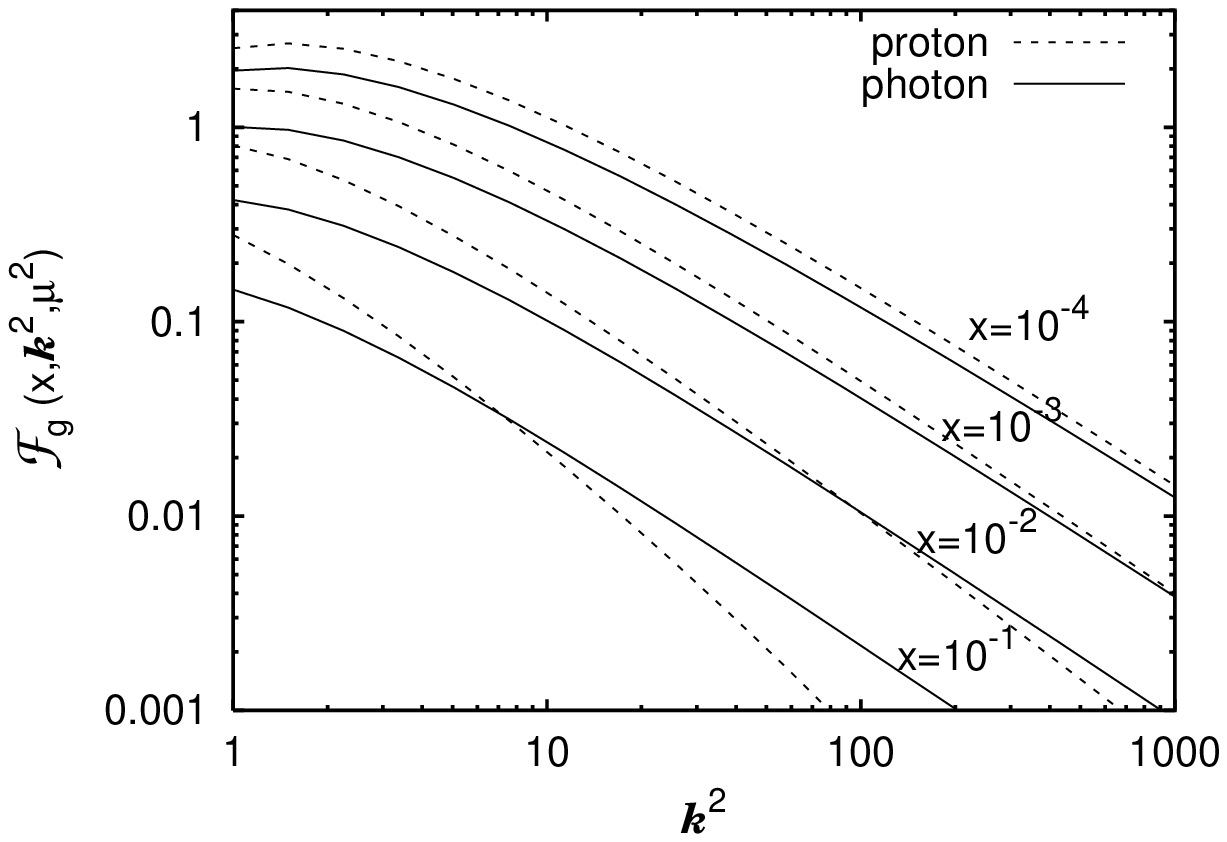} \\
a) & b) \\
\end{tabular}
\caption{Comparison between the off-shell gluon distributions in the proton and
in the photon, using KMR approach, as a function of (a) longitudinal momentum
fraction $x$ and (b) transverse momenta $\tdm k$ of the gluon, for
fixed values of $\tdm k^2$ (GeV$^2$) and $x$, respectively.}
\label{updf}
\end{center}
\end{figure}


We have extended the KMR formalism to the case of the photon.
In the following we will use the unintegrated gluon density in the photon
defined by equation~(\ref{kmreq}).
The conventional gluon and quark densities in the photon are expressed 
following the Gl\"uck-Reya-Schienbein (GRS) parameterization \cite{GRS-photon}.
Thus, $g(x,Q^2)$ and $q(x,Q^2)$ consist of two components, a
pointlike (perturbative) component parameterized in \cite{GRS-photon} and a
hadronic component, given by the parton distribution functions in the pion
\cite{GRS-pion}. For instance, for the gluon we use
\be
x g(x,Q^2) = \frac{1}{\alpha_{em}} x g_{pl}(x,Q^2) + (G_{\omega}^2 + G_{\rho}^2)
  x g_{\pi}(x,Q^2) ,
\label{grs}
\ee
with $G_{\omega}^2=0.043$ and $G_{\rho}^2=0.50$, while the respective formulae
for quarks can be found in \cite{GRS-photon}. 
The obtained unintegrated gluon density is defined for 
$\tdm k^2 > k_{0}^2 = 0.5$ GeV$^2$, which is the starting scale for
the GRS distribution. However, an extrapolation to cover the whole
range in $\tdm k^2$ has been performed,
extending the gluon density to values of $\tdm k^2 <  k_{0}^2
$by normalizing it to the GRS distribution in the following way
${\cal F}_g(x,\tdm k^2,\mu^2) = xg(x, k_{0}^2)/ k_{0}^2$.

Another solution \cite{CCFM-proton} to the CCFM equation was found using the
"single loop" approximation, when small-$x$ effects can be neglected in the CCFM
equation for medium and large $x$. Thus an exact analytic solution can be
obtained, and a comparison between this analytic solution for the proton and the
KMR approximation shows quite good agreement \cite{CCFM-proton}. Similarly, 
for the photon \cite{CCFM-photon}, the unintegrated gluon distributions 
obtained from the exact solution of the CCFM equation in the single loop
approximation can be well represented by the KMR distributions
constructed using the integrated quark and gluon distributions and the Sudakov form factor.

Although the KMR constructions of unintegrated gluon distributions for the photon
and proton are similar, the distribution in the photon is significantly
different due to the pointlike component. As can be seen in Fig.~\ref{updf},
a direct comparison between the unintegrated gluon in the photon and in the
proton shows a relative enhancement for large $x$ and 
large $\tdm k$ in the case of the  photon.  
This enhancement is due to gluon emissions from the perturbative
quark box, making the gluon distribution much harder as compared 
to the proton for large values of $x$. For small values of $x$, 
the similar shape of both distributions indicates that the 
information about the shapes, contained in the input at large~$x$, 
is partially lost in evolution. Note that for the respective KMR 
gluon density in the proton we have used the conventional 
GRV parameterization~\cite{GRV} for the proton.

\subsection{The GBW gluon}
\label{sec2b}

Another parameterization of the unintegrated gluon density in the photon
can be obtained using a simple generalization of the 
Golec-Biernat and W\"usthoff (GBW) parameterization of the gluon in the proton
\cite{GBW}. The unintegrated gluon density introduced in \cite{GBW}
for the proton
\be
\label{GBWgluon}
{\cal F}_g(x,\tdm k^2)=\frac{3\sigma_0}{4\pi^2\alpha_s}
R_0^2(x) \tdm k^2 \exp(-R_0^2(x)\tdm k^2) ,
\ee
is related to the effective dipole cross section within the saturation model
\be
\label{sigsat}
\hat \sigma(x,r^2)=\sigma_0 [1-\exp(-r^2/(4R_0^2(x)))],
\ee
which describes the interaction between a proton and a color $q\bar{q}$ dipole
coming from a photon fluctuation
\be
\label{gammap}
\sigma ^{\gamma^*p} = \sum_{a=1}^{N_f} \int_0^1 dz \int d^2{\tdm r}  \; 
|\Psi^a (z,{\tdm r})|^2 \, \hat \sigma(x,r^2) .
\ee
In the above equations, $\tdm r$ denotes the transverse separation of the quarks
and $z$ gives the longitudinal momentum of the quark in the photon. The
wave function of the photon is represented by $\Psi^a (z,{\tdm r})$,
where $a$ indexes the flavor of the quark in the dipole, 
and its form can be found in \cite{GBW}, for instance.
Thus, for real photons the mass of quark $a$ gives the characteristic
scale for the dipole distribution in the photon.
The saturation radius is given by
\be
\label{r0}
R_0(x)=\frac{1}{Q_0} \left(\frac{x}{x_0}\right)^{\lambda/2} ,
\ee
with $\sigma_0=29.12$ mb, $\alpha_s=0.2$, $Q_0=1$ GeV, $\lambda=0.277$ and 
$x_0=0.41 \times 10^{-4}$. The three free parameters $\sigma_0$, $\lambda$ and
$x_0$ have been fitted and the saturation model describes successfully both
inclusive and diffractive $\gamma^* p$ scattering \cite{GBW}.
The mass of the light quarks $u$, $d$ and $s$, $m_q = 0.21$~GeV 
is taken from a fit of the saturation model to inclusive two-photon
observables~\cite{TKM}.

Following the generalization of the GBW saturation model for the case of 
$\gamma^* \gamma^*$ scattering introduced in
\cite{TKM}, one can easily construct in a 
similar fashion the unintegrated gluon distribution in the photon. In such 
a case, we consider the scattering of two color dipoles, into which the photons
fluctuate, one light $q\bar{q}$ dipole and one heavy $Q\bar{Q}$ dipole
\be
\label{sigmaTKM}
\sigma^{\gamma^*\gamma^*}=
\sum_{a=1}^{N_f} \int_0^1dz_1\int d^2 {\tdm r_1}|\Psi^a(z_1,{\tdm r_1})|^2
\int_0^1 dz_2\int d^2 {\tdm r_2}|\Psi^Q(z_2,{\tdm r_2})|^2 
\; \sigma^{dd}_{a}(\bar x,r_1,r_2) .
\ee
The heavier dipole with the transverse separation $\tdm r_2$ provides
the hard scale at which the dipole content of the second photon is probed.
In this configuration, the relative size of the heavy dipole 
($\langle r_2 \rangle \sim 1/2m_Q$) is smaller than that of the light dipole
($\langle r_1 \rangle \sim 1/2m_q$).
The effective  dipole-dipole cross section $\sigma^{dd}(\bar x,r_1,r_2)$ is a 
generalization of the GBW cross section from eq.~(\ref{sigsat}), introducing
an effective dipole separation radius $\tdm r_{eff}$ depending on the
size of the two dipoles \cite{TKM}. In our configuration with 
one heavy dipole  $r_{2}^2 \ll r_1^2$, the effective cross section reduces
to eq.~(\ref{sigsat}), depending only on $\tdm r_2$.

\begin{figure}[th]
\begin{center}
\epsfig{width= 0.5\columnwidth,file=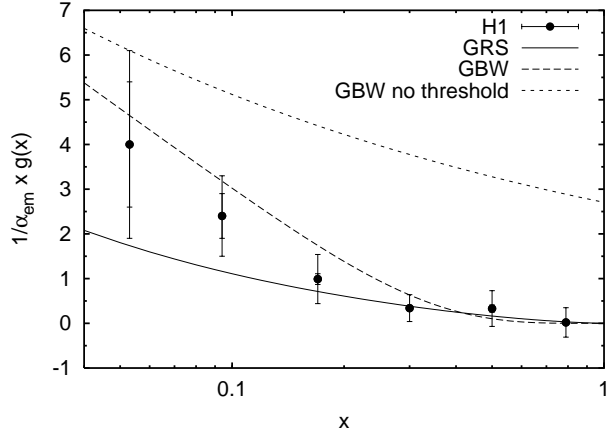,clip=}
\caption{The gluon distribution $xg(x)$ in the photon as a function of $x$.
The data points describe the gluon content in the photon extracted
from the photoproduction of hard dijets (mean $p_t^2=74$ GeV$^2$) observed at
HERA \cite{H1dijets}. The GRS parameterization \cite{GRS-photon} is compared with
the integrated GBW distribution from (\ref{TKMgluon})
at $\mu^2 = p_t^2$, with and without the 
phenomenological threshold factor included.}
\label{dijet}
\end{center}
\end{figure}


In this region, the integrals from the equation~(\ref{sigmaTKM})
over $z_1$ and $\tdm r_1$ of the first dipole can be performed
independently. In the leading logarithm $\log(m_Q/m_q)$ approximation, 
the result of this integration is dominated by the contribution
\be
\label{nodipoles}
N_{d}(\mu) = 
\sum_{a=1}^{N_f} \int_0^1dz_1 \int_{1/\mu^2}^{\infty} d^2 {\tdm r_1} 
|\Psi^a(z_1,{\tdm r_1})|^2 ,
\ee
with a lower cut-off in $\tdm r_1$ provided by the typical
size of the  heavy dipole, $1/\mu \sim 1/2m_Q$. This integral may 
be interpreted  as the number of dipoles in the photon at the scale given
by the mass of the heavy quark. The final result after the integration will be
a form for the $\gamma^* \gamma^*$ cross section which is similar
to the $\gamma^* p$ cross section in eq.~(\ref{gammap})
\be
\label{newsigma}
\sigma^{\gamma^*\gamma^*}=
N_{d}(2m_Q) \times
\int_0^1 dz_2\int d^2 {\tdm r_2}|\Psi^Q(z_2,{\tdm r_2})|^2 
\; \hat \sigma (x,r_2) .
\ee
The number of  dipoles $N_{d}$ in the photon is found to be
1.46$\cdot \alpha_{em}$ for charm production ($M_c=1.3
$GeV) and 2.43$\cdot \alpha_{em}$ for bottom production ($M_b=4.8$ GeV).

\begin{figure}[th]
\begin{center}
\begin{tabular}{cc}
\epsfig{width= 0.45\columnwidth,file=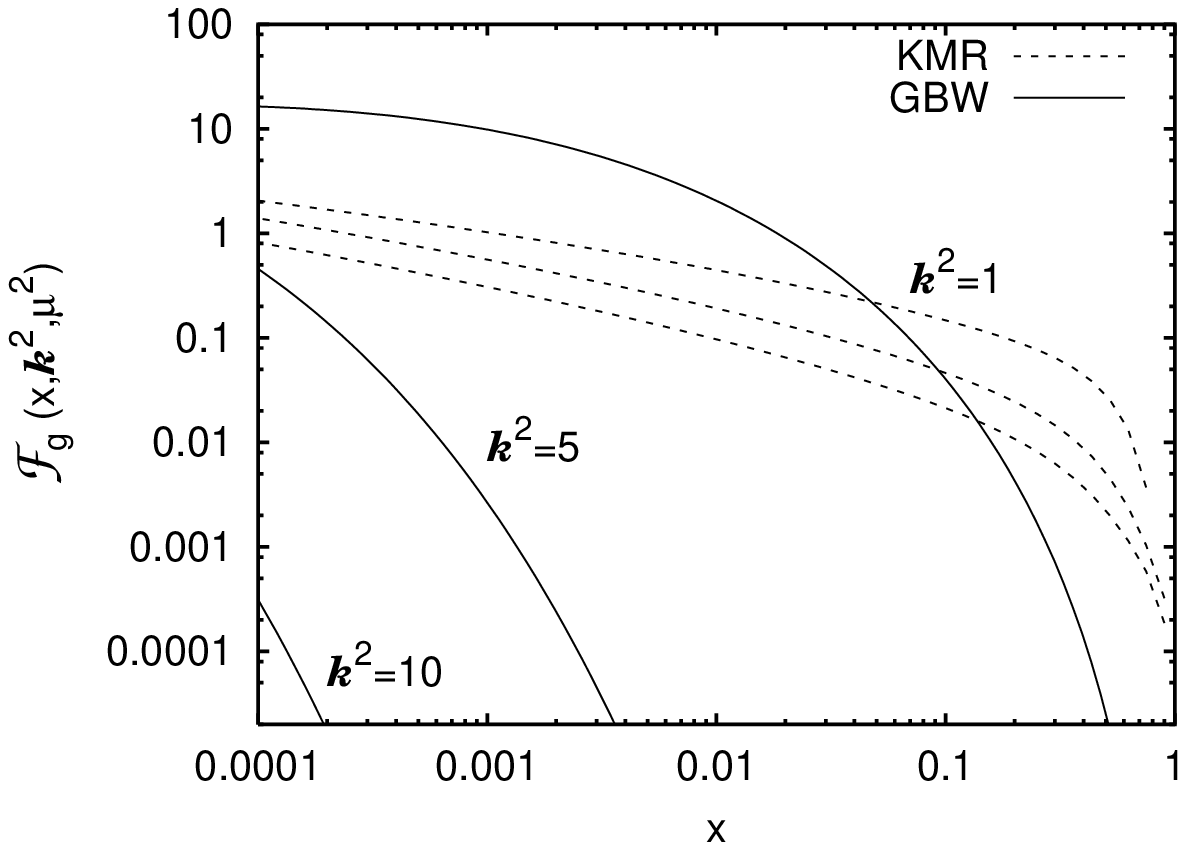} &  
\epsfig{width= 0.45\columnwidth,file=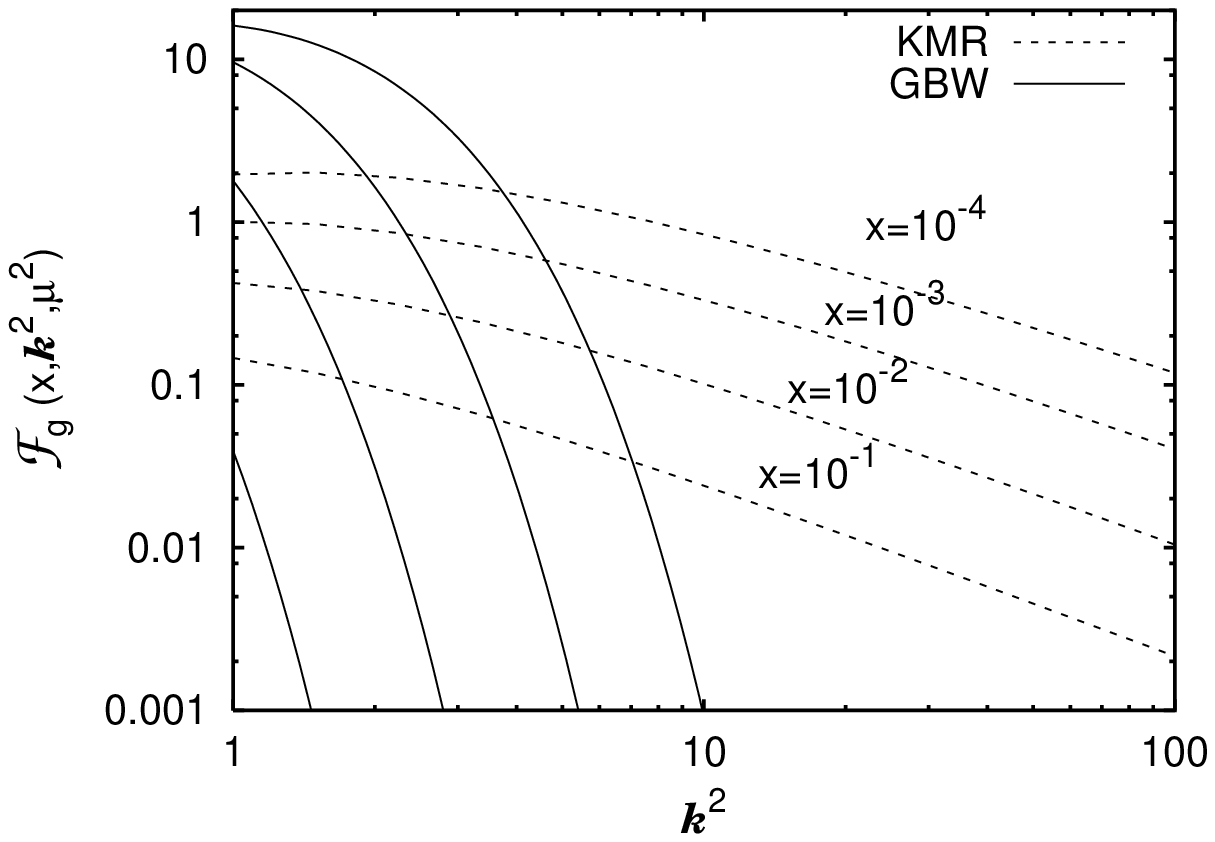} \\
a) & b) \\
\end{tabular}
\caption{Comparison between the unintegrated gluon distributions in the 
photon, using the KMR approach (dashed lines) and
the generalized GBW parameterization (solid lines)
as a function of (a) longitudinal momentum
fraction $x$ and (b) transverse momenta $\tdm k$ of the gluon, for
fixed values of $\tdm k^2$ (GeV$^2$) and $x$, respectively.}
\label{kmr-gbw}
\end{center}
\end{figure}


The extraction of the gluon density in the photon from eq.~(\ref{newsigma})
will give a parameterization similar to the one in eq.~(\ref{GBWgluon})
\be
\label{TKMgluon}
{\cal F}_g(x,\tdm k^2,\mu^2)=
N_d(\mu) \times \frac{3\tilde \sigma_0}{4\pi^2\alpha_s}
R_0^2(x) \tdm k^2 \exp(-R_0^2(x)\tdm k^2) .
\ee
This includes a multiplicative factor given by the number of dipoles
in the photon, and the parameter $\tilde \sigma_0 = \frac23 \sigma_0
$as introduced in the generalization of the GBW model \cite{TKM} 
(the factor $2/3$ is a reminder of the quark counting rule, with
$\tilde \sigma_0$ representing the cross section in the blackness limit 
for the photon, and respectively, $\sigma_0$ for the proton).
The number of dipoles that enters in the gluon
density has an intrinsic dependence on the hard scale given by the heavy quark
mass, which propagates as a secondary scale at the level of the unintegrated
gluon distribution. All other parameters from the original GBW parameterization
are kept unchanged.

As described in \cite{TKM}, in order to extend the color dipole model
to moderate and large $x$ values, the introduction of phenomenologically motivated
threshold factors was necessary.  Thus,  we have imposed the following form
on the total cross section in $\gamma\gamma$ interactions, $\sigma^{\gamma^*\gamma^*} \sim (1-x)^{2n_{spect}-1}$. In the $k_t$ factorization approach, this factor can be understood
as having a dual contribution, from 
the off-shell matrix elements and the unintegrated
gluon distributions, which consider the correct kinematics of the hard
process. To account for the full kinematics in the unintegrated
density of the photon, we will introduce such a multiplicative
factor in this density. 
When probing the gluon content of a hadron with a photon, 
only sea quarks can be picked, so the number of spectators is $n_{spect}=4
$in the case of a proton (3 constituent quarks plus 1 sea quark),
and $n_{spect}=3$ for the photon case (2 quarks from the dipole plus 1 
from the sea).

Multiplying ${\cal F}_g(x,\tdm k^2,\mu^2)$  from (\ref{TKMgluon}) 
with the factor $(1-x)^5$ and integrating it
to obtain the corresponding on-shell gluon density,
we find the same dependence with $x$ as the GRS distribution for large-$x$.
As seen in Fig.~\ref{dijet}, the integrated gluon distribution
provides a better agreement with the existing data extracted
from photoproduction of hard dijets at HERA~\cite{H1dijets}, as compared
to the case where no threshold factor was included.
Similarly, we have extended the applicability of the GBW gluon distribution
for the proton (\ref{GBWgluon}) for large values of $x$ 
by introducing the multiplicative factor $(1-x)^7$.

The above unintegrated gluon distribution obtained using the 
extended saturation model exhibits different $x$ and $\tdm k
$dependence than the previous density stemming from the KMR approach.
These differences are best expressed in Fig.~\ref{kmr-gbw}, where
a suppression for large values of $\tdm k$ and an enhancement at
small momenta can be seen in the GBW gluon. In spite of their
unlikeness, the two distributions give quite similar results when
integrated over the transverse momenta. Figure~\ref{dijet} shows
how the integrated KMR distribution, similar to the conventional
gluon density GRS, and the integrated GBW gluon compare with data.
Let us note how poorly constraining data is for the gluon content
of the photon, and that a new fit could give an increase 
in the gluon distribution which can alter significantly predictions
for cross sections based on these distributions.

\section{Cross sections for heavy quark production}
\label{sec3}

\begin{figure}[ht]
\begin{center}
\begin{tabular}{ccc}
\epsfig{width= 0.3\columnwidth,file=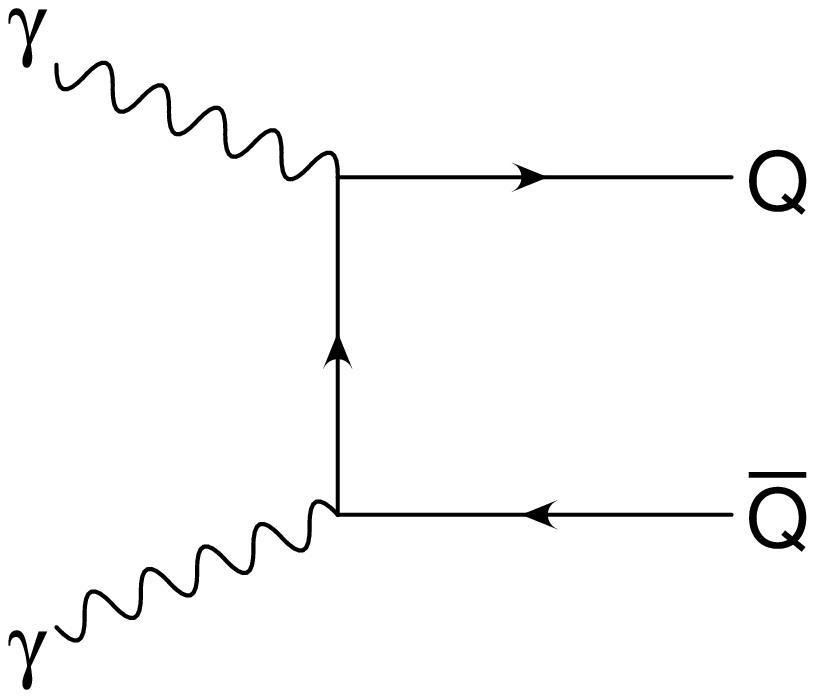,clip=} &
\epsfig{width= 0.3\columnwidth,file=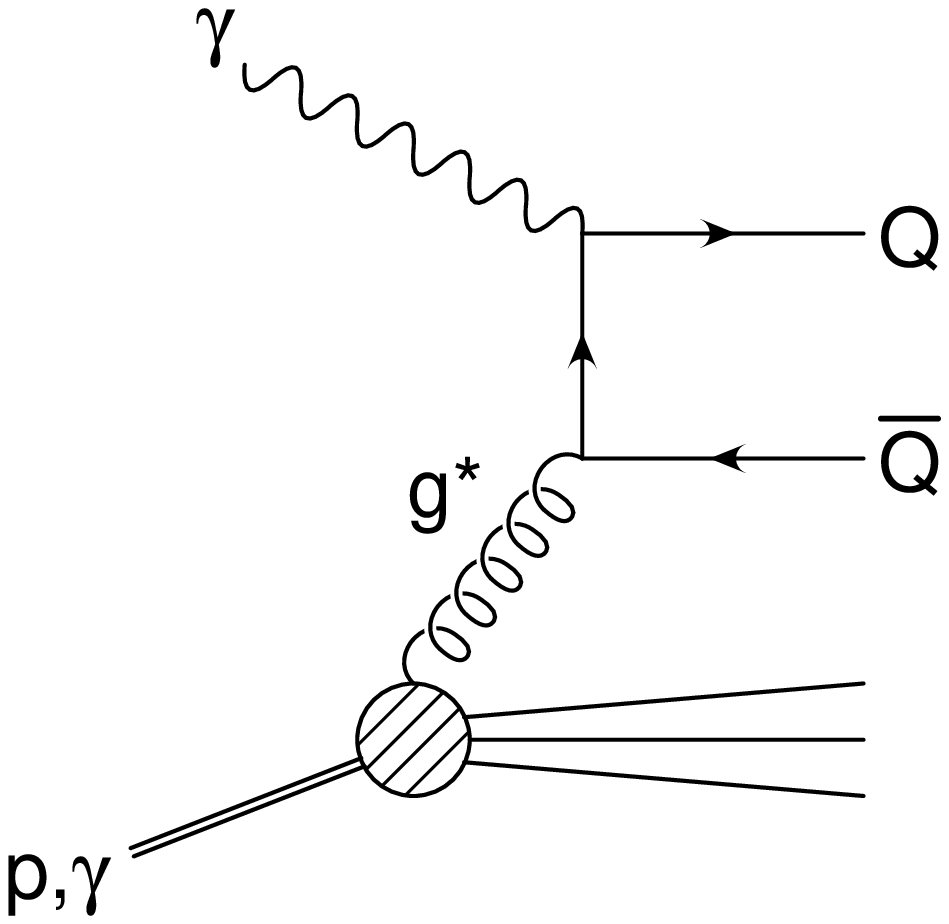,clip=} &
\epsfig{width= 0.3\columnwidth,file=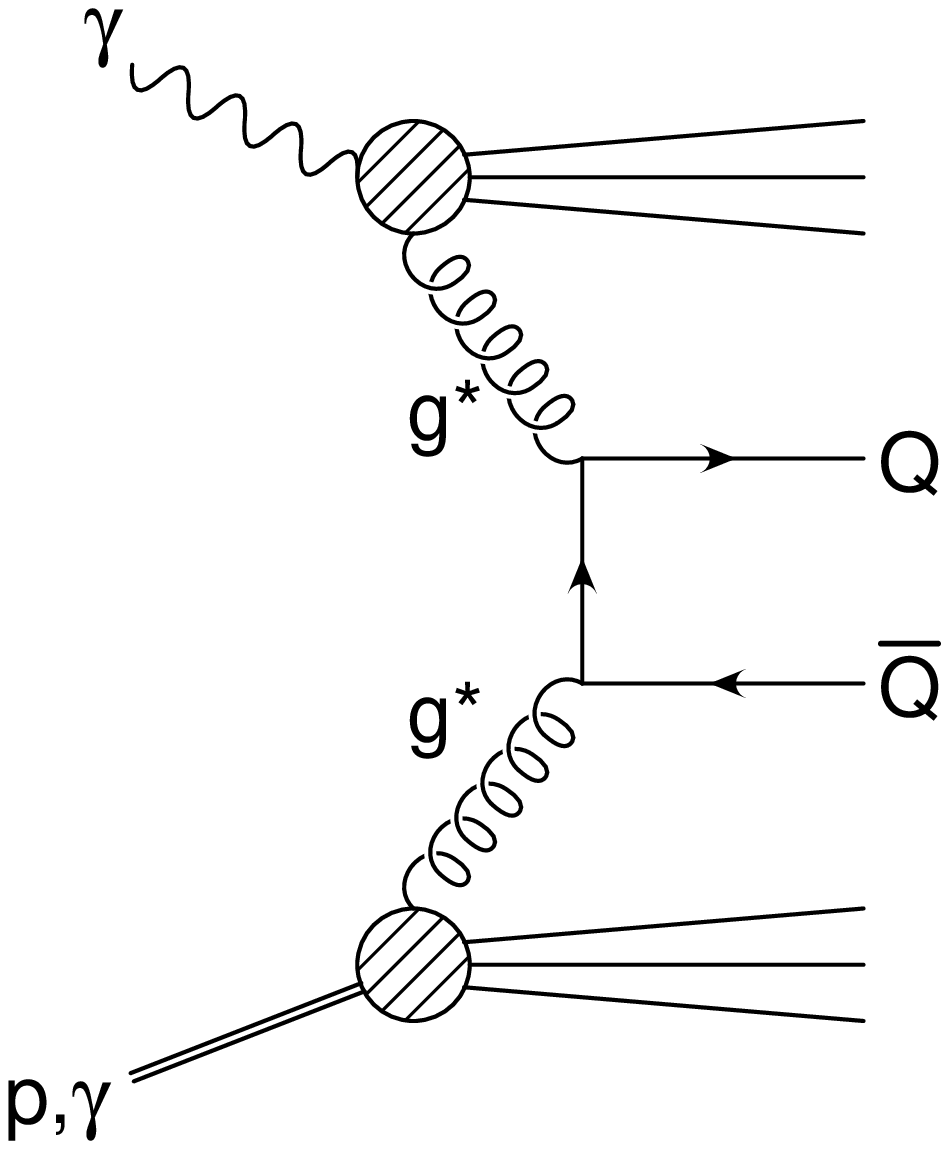,clip=} \\
a) & b) & c) \\
\end{tabular}
\caption{Diagrams illustrating the heavy quark production in $\gamma\gamma
$or $\gamma p$ collisions through  different mechanisms: a) direct production
(only for $\gamma\gamma$ interactions), b) photoproduction in single resolved
case, where one photon or the proton is resolved,  and c) double resolved case,
where both incoming particles are resolved.}
\label{diagrams}
\end{center}
\end{figure}

Heavy quarks may be produced in two-photon collisions
by one of the three mechanisms: a direct production
(Fig.~\ref{diagrams}a), a photoproduction off a resolved 
photon (Fig.~\ref{diagrams}b) and a by a double resolved process
(Fig.~\ref{diagrams}c). The direct contribution to the process
$\gamma\gamma \to Q\bar Q X$ is governed by elementary 
QED amplitudes\cite{Budnev}. In the case of proton-photon
scattering, this component is absent.

In the collinear limit one uses the leading twist term of the
operator product expansion, neglecting the transverse momenta of the partons.
The cross section for heavy quark 
photoproduction off one of the photons being resolved 
reads
\be
\sigma^{cf} _{1R} (s,M^2) = \sum_{i} 
\int {dx \over x} f_i (x,\mu^2)  
\hat\sigma^{cf} _{\gamma i} (M^2, xs), 
\label{gpcl1r}
\ee
where the partonic cross sections $\hat\sigma^{cf}_{\gamma i}$ are
well known up to the NLO approximation and $f_i(x,\mu^2)
$is the parton~$i$ distribution function in the photon,
at the factorization scale $\mu \simeq 2 M$.  
In the above, $s$ denotes the $\gamma\gamma$ invariant mass 
squared, and $M$ is the quark mass.

In the $k_t$~factorization formalism, the complete kinematics
of the gluon-photon fusion is taken into account and 
the small light-cone component of the longitudinal momentum 
of the gluon is integrated out. Then, the cross section 
takes the following form 
\be
\sigma^{kf}_{1R} (s,M^2) = \sum_{i} 
\int {dx \over x}  {d^2\tdm k \over \pi}   
{\cal F}_i (x, \tdm k^2,\mu^2)\, 
\hat \sigma^{kf} _{\gamma i} (\tdm k^2, M^2, x s), 
\label{gpkt1r}
\ee
in which the unintegrated parton density 
${\cal F}_i (x, \tdm k^2,\mu^2)$ and the off-shell
partonic cross section $\hat \sigma^{kf} _{\gamma i} (\tdm k^2, M^2, x_1 x_2 s)$ 
are employed. 
The partonic cross sections are evaluated using off-shell matrix 
elements. The form of $\hat\sigma^{kf}$ is well known in the 
literature, see for example \cite{CCH}, and we quote it in Appendix~\ref{apB}. 
Formulae (\ref{gpcl1r}) and (\ref{gpkt1r}), with the appropriate
substitution of parton densities, are valid also for the 
photoproduction off the proton. It is important to
note that, in a $\gamma\gamma$ collision one or the other 
photon may be resolved, thus the inclusive cross section for heavy 
quark production acquires an additional factor of~2.
At the LO approximation, only gluons contribute to 
the $Q\bar Q$ production.

The double-resolved contribution to the process
$\gamma\gamma\to Q\bar Q X$ is described by
\be
\sigma^{cf}_{2R} (s,M^2) = \sum_{i,j} 
\int {dx_1 \over x_1} {dx_2 \over x_2} 
{f}_i (x_1,\mu^2) {f}_j (x_2,\mu^2)
\hat \sigma^{cf} _{ij} (M^2, x_1 x_2 s), 
\label{gpcf2r}
\ee
in the collinear limit and the cross section
in the $k_t$ factorization formalism reads
\be
\sigma^{kf}_{2R} (s,M^2) = \sum_{i,j} 
\int {dx_1 \over x_1} {dx_2 \over x_2} 
{ d^2\tdm k_1 \over \pi}  { d^2\tdm k_2 \over \pi}  
{\cal F}_i (x_1, \tdm k_1^2,\mu^2)\, {\cal F}_j (x_2, \tdm k_2^2,\mu^2)\,
\hat \sigma^{kf} _{ij} (\tdm k_1, \tdm k_2, M^2, x_1 x_2 s), 
\label{gpkt2r}
\ee
where $\hat \sigma^{kf} _{ij}$ for gluons is given in Appendix~\ref{apA}
(following from \cite{BE}).
Analogously, one of the photons may be replaced by
the proton in order to obtain a resolved photon contribution 
to heavy quark photoproduction off the proton. 

In the following, we shall restrict ourselves
to the effects of transverse momentum in the gluon
kinematics, and the quark contribution 
will only be taken in the collinear approximation.
This should not affect significantly the results,
as the heavy quark production is driven mostly by
exchanges of gluons. For the collinear limit, all the
theoretical estimates were obtained using the
PYTHIA Monte Carlo \cite{PYTHIA}.

\section{Results for heavy quark production}
\label{sec4}

The measurements of cross sections for the inclusive charm and bottom
production in $e^+ e^-$ and $e p$ collisions
were performed at LEP and HERA respectively.
For bottom production at LEP, only the total rate
$e^+e^- \to e^+ e^- b\bar b X$ was measured~\cite{LEPHQ}, representing
an average of the $\gamma\gamma \to  b\bar b X
$cross section weighted with the flux of photons in the
electrons. For charm, the cross section
$\sigma(\gamma\gamma \to c\bar c X)$ was determined
for different $\gamma\gamma$ collision energies~\cite{LEPCC}.
At HERA, the cross section for $\gamma p \to b\bar b X
$is known~\cite{HERABB} for the collision energy averaged between
$W= 94$~GeV and $W = 266$~GeV. The data for charm production
at HERA~\cite{exp_f2c} are available for virtual photons, at different
virtualities $Q^2$ and collision energies $W$ in the form
of $F_2 ^{\mathrm charm}(x,Q^2)$. We give in this
section the theoretical estimates for these processes based on the
$k_t$ factorization formalism and study the theoretical uncertainties.

\subsection{Theoretical uncertainties}
\label{sec4a}

The cross sections for heavy quark production are
described by formulae (\ref{gpkt1r}) and (\ref{gpkt2r})
with the LO~partonic cross sections given by (\ref{gamgluqq})
and (\ref{glugluqq}). The results of the numerical evaluation
of these formulae depend somewhat on the model for
unintegrated parton densities and the choice of
parameters. We have examined the following options for
different elements of the model:

\paragraph{The heavy quark mass $M$}is plagued by
a fundamental uncertainty due to confinement of color
-- there are no free quarks, and consequently,
there is no on-shell quark mass. The running quark mass
in QCD varies with the scale. It is not clear, at which
scale the quark masses in the matrix elements should
be evaluated, because the virtualities of heavy
quarks are different for different lines in the
relevant Feynman diagrams. Thus, we shall assume
for the $b$~quark that $4.5$~GeV~$< M_b < 5$~GeV and for
the $c$~quark that $1.3$~GeV~$< M_c < 1.5$~GeV.

\paragraph{The energy scale $\bar\mu$}that enters
the running formula of the strong coupling constant
$\alpha_s(\bar\mu^2)$ in the partonic cross section (see Appendix)
is usually chosen to be of order of the typical momentum transfer 
characterizing the process. 
However, the optimal value of
$\bar\mu$ is such, that the contribution of higher
orders in the perturbative expansion is minimal.
Thus, without knowledge of higher order corrections,
$\bar\mu$ is uncertain and, in order to account
for this ambiguity we considered three options:
(1)~$\bar\mu^2 = M^2 + \tdm p^2$ (standard),
(2)~$\bar\mu^2 = (M^2 + \tdm p^2)/4$ (low scale)
and (3)~$\bar\mu^2 =4 M^2$ (large scale), where
$\tdm p = \tdm k$ (gluon transverse momentum) for
the gluon-photon fusion (\ref{gamgluqq})
and $\tdm p = \tdm k_1 - \tdm k_2$ for the two-gluon
process (\ref{glugluqq}).

\paragraph{Running of the coupling constant.}
We use the standard one-loop running formula
for $\alpha_s$, with four flavors. We use, as
a default choice, $\Lambda_{\mathrm QCD}=140$~MeV,
such that $\alpha_s(M_Z^2)=0.117$ as given by the latest
QCD fits. We  also test the PYTHIA default value:
$\Lambda_{\mathrm QCD}=250$~MeV, corresponding to the value of
$\alpha_s(M_Z^2)=0.128$.

\paragraph{Unintegrated parton distributions.}
For the proton, we take into account the CCFM gluon distribution,
as given in the CASCADE Monte Carlo \cite{CASCADE}, 
the unintegrated gluon obtained
from the GRV and MRST parameterizations using the KMR method (uGRV and
uMRST) and the gluon following from the saturation
model (GBW). In the case of a real photon, the KMR method
is applied to the GRS gluon (uGRS) and an alternative
model of the unintegrated gluon is given by the saturation
model for photons (GBW$\gamma$), as explained in Sec.~\ref{sec2a}.
Furthermore, we vary the factorization scale in the two-scale
gluon distribution ${\cal F}_g(x,\tdm k^2,\mu^2)$ between $\mu = M
$and $\mu = 2M$.

\paragraph{Parton momentum fraction $x$.} In the collinear
approximation, one neglects the non-zero transverse momentum
$\tdm k$ of the incoming parton whereas in the $k_t$ factorization
approach,  the transverse momentum is included in the kinematics
of partonic scattering. Thus, for the virtual photon-gluon fusion,
the invariant mass of the system is
$\hat s_ {\gamma g} = x s$ in the collinear approximation and
$\hat s'_ {\gamma g} = x s - \tdm k^2$ when the complete kinematics
are taken into account. 
In a standard approximation method, the cross section in the collinear
factorization can be obtained from the one in the $k_t$ factorization, 
using a substitution 
$\hat\sigma^{kf}(xs,\tdm k^2, Q^2) \rightarrow 
\hat\sigma^{kf}(xs,\tdm k^2 = 0, Q^2) \theta(Q^2-\tdm k^2)$,
and the integrated parton distributions as defined by the eq.~(\ref{xgx}).
However, such approximation neglects the fact that the kinematical
threshold in DIS for $\hat\sigma^{kf}$ depends on $\tdm  k^2$,
{\em i.e.} 
the kinematical threshold
for the virtual photoproduction is located at $x \simeq Q^2 /s$ in the
collinear approximation and at
\be
\bar x (\tdm k^2) \simeq (Q^2 + \tdm k^2) / s \simeq x [1+ \tdm k^2 / Q^2]
\label{xbar}
\ee
in the $k_t$ factorization framework. 

In order to investigate the inclusion of the latter threshold treatment,
the standard approximation could suffer modifications by introducing
a rescaled variable $z=x Q^2/(Q^2 + \tdm k^2)$. This will improve
the approximation and lead to an alternative relation between the 
$k_t$ factorization and collinear approximation expressions, where
\be
z g(z,Q^2) = \int_0 ^{Q^2} d\tdm k^2\; 
{\cal F}_g( z (Q^2 + \tdm k^2)/Q^2,\tdm k^2,Q^2).
\label{xinfg}
\ee
Thus, the use of the rescaled variable is a way to quantify the ambiguity
in obtaining the unintegrated gluon distribution from the integrated one.

In our investigation of the effects of the threshold treatment in the
evaluation of the heavy quark production cross section, we consider
a substitution
\be
x' = {x \over  1 + \tdm k^2 / (4M^2)}
\label{xresc}
\ee
in the unitegrated gluon distributions ${\cal F}_g (x',\tdm k^2,\mu^2)$.
Note, that for massive quark photoproduction,
$4M^2$ replaces $Q^2$. Such rescaling is not necessary
for the CCFM gluon, where the kinematical effects of the transverse
momentum are already accounted for in the $F_2$ fits.

\subsection{Results for $\gamma p$ interactions}
\label{sec4b}

\begin{figure}[ht]
\begin{center}
\begin{tabular}{cc}
\raisebox{40ex}[0cm][0cm]{a)} & 
\epsfig{width= 0.5\columnwidth,file=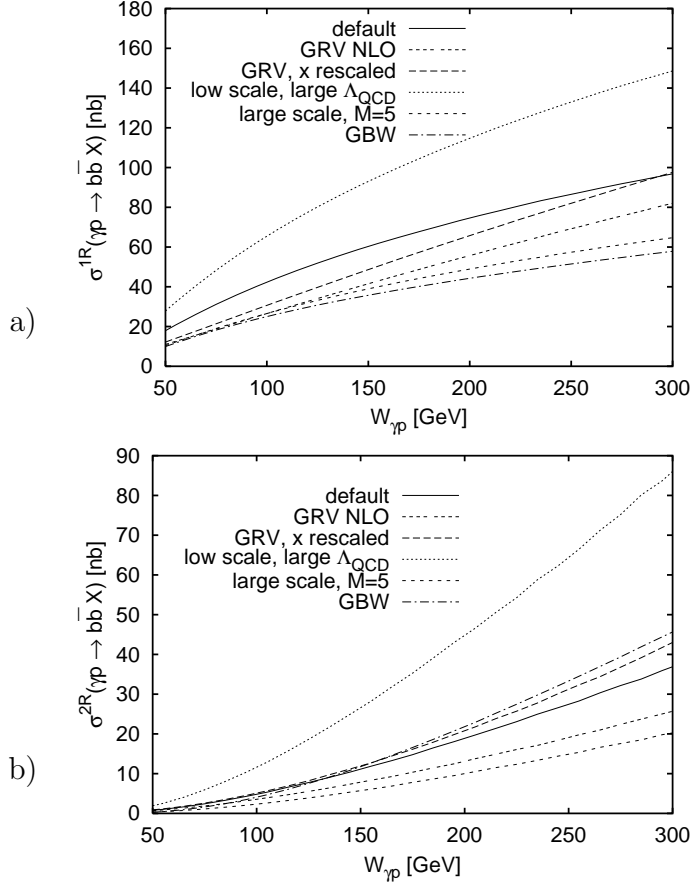} \\
\raisebox{10ex}[0cm][0cm]{b)} & 
\end{tabular} 
\caption{Cross sections of \bbbar production in $\gamma p$ interactions
from the $k_t$ factorization approach, showing a) the single resolved
contribution and b) the double resolved contribution. Details of the
curves are presented in the text.
}\vspace*{-2mm}
\label{gpbb}
\end{center}
\end{figure}

In Fig.~\ref{gpbb} we give a set of results for cross sections
for $\gamma p \to b\bar b X$ with the direct photon (Fig.~\ref{gpbb}a)
and with the resolved photon (Fig.~\ref{gpbb}b).
The default results (full curves)
are  obtained by taking the CCFM unintegrated gluon in the proton
(from CASCADE), $M_b = 4.5$~GeV, $\mu = 10$~GeV,
$\Lambda_{\mathrm QCD}=140$~MeV and the {\em standard}
scale for the running coupling (see Sec.~\ref{sec4a}). For the
gluon in the photon, the KMR method is applied to obtain the
unintegrated GRS parameterization (uGRS).
Besides the default choice, we also use the unintegrated GRV
distribution (uGRV), based on the GRV~NLO parameterization,
and the GBW gluon\footnote{In the definition of the unintegrated gluon
in the saturation model, one assumes a fixed $\alpha_s = 0.2$,
and, consequently, the same choice was made for
the coupling in matrix elements when GBW gluon was used.},
leaving other parameters unchanged.
Furthermore, we investigate the impact of the $x$-rescaling,
for the uGRV gluon (direct photon) and for both  the uGRV and
uGRS gluon (resolved photon).
We modify the default choice of the QCD parameters:
we use $\Lambda_{\mathrm QCD}=250$~MeV and the {\em low}
scale for running, in order to obtain an upper limit for the cross section.
A very conservative estimate follows from the choice
$M_b=5$~GeV and the {\em large} energy scale for $\alpha_s$.
We have checked, that using the unintegrated MRST gluon
and variations of the renormalization scale $\mu$ have a minor
influence on the results, hence we do not include the
corresponding curves in the figure.

\begin{figure}[ht]
\begin{center}
\begin{tabular}{cc}
\multicolumn{2}{c}{\epsfig{width= 0.95\columnwidth,file=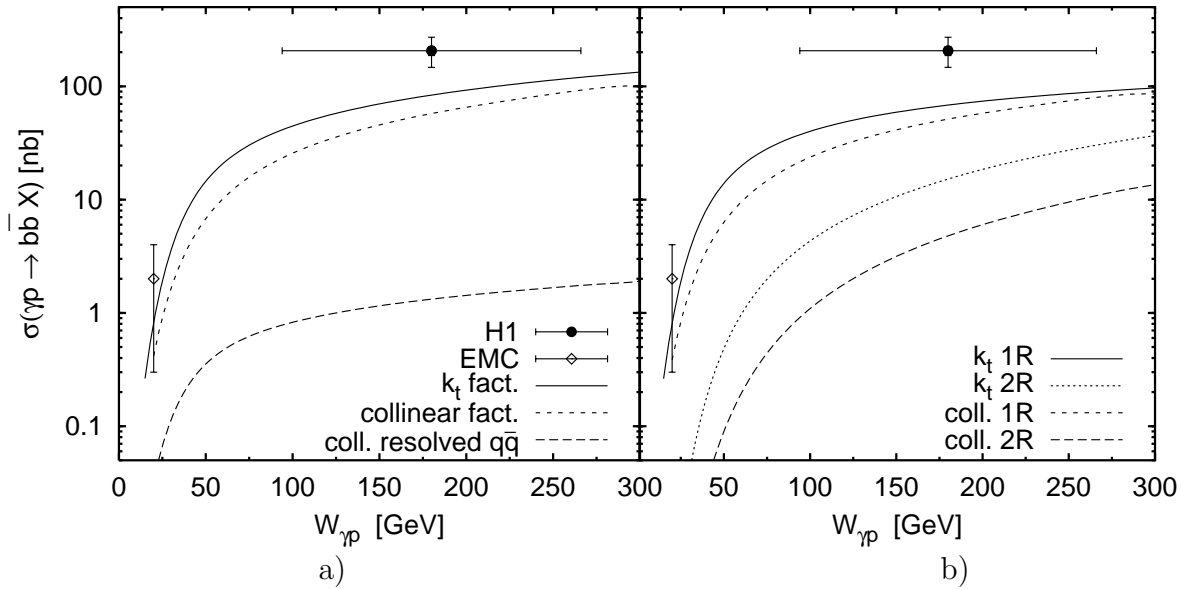}}\\
 \hspace{10em} a) & \hspace{10em}  b)
\end{tabular} 
\caption{Comparison between the collinear factorization and the $k_t
$factorization approaches to \bbbar production in $\gamma p$ interactions.
Shown are the experimental data from EMC~\cite{EMCBB} and H1~\cite{HERABB}.
a)~The total \bbbar cross section in the $k_t$ factorization
(solid line) includes single and double resolved contributions.
In the collinear factorization, the total cross section (short dashed)
also includes  the specific contribution 
from the $q\bar q$ annihilation for the double resolved case
(long dashed).
b) Comparison between the single resolved components in the
$k_t$ (solid line) and collinear (short dashed) factorizations,
and between the double resolved components in the $k_t$ (dotted)
and collinear (long dashed) factorizations.
}\vspace*{-4mm}
\label{gpbb-coll}
\end{center}
\end{figure}

It is visible in the figure, that the contribution from
the resolved photon is significant -- roughly 20--30\% of
the direct photon cross section for the default choice
of parameters. We have checked, that in all cases, the resolved
photon contribution rises faster with the energy $W$ than
the direct photon one. The results are rather stable against
modifications of the unintegrated gluon and the quark mass.
The largest contribution to the uncertainty 
of the cross section comes from the
details of the QCD running of $\alpha_s$ and the choice of
energy scale -- reflecting a potential influence of higher
order corrections. When the {\em low} scale of $\alpha_s
$and the large value of $\Lambda_{\mathrm QCD}$ are
assumed, the direct contribution gets enhanced by about
50\% and the resolved one gets doubled. In this extreme
case, the calculation results for the sum of the direct and
resolved photon cross sections 
$\sigma(W = 180$~GeV)$ = 143$~nb 
is close to the experimental data point 
$\sigma = 206 \pm 19 ^{+46 } _{-40}$~nb (see Fig.~\ref{gpbb-coll}).
Thus, although the QCD predictions are significantly lower 
than the data, inconsistency cannot be claimed, due
to the large experimental error and the large theoretical
uncertainty coming from uncontrolled higher order corrections.

The saturation model estimate of the direct photon
contribution to bottom production is lower than the
default result. The double resolved contribution obtained using
the GBW parameterization for both the photon and the proton 
is slightly larger than its KMR counterpart.
The constraint we have imposed on the parton density
at large $x$ via the threshold factor plays a very 
important role for the resolved photon case, being dependent
on the gluon at relatively large~$x$.
The consistency of the standard QCD and saturation model 
results is not surprising, as both sets of parameterizations
are constrained by the same experimental data (see Fig.~\ref{dijet}).
It is, however, striking that the unintegrated GBW gluon
leads to a lower cross section than the uGRS gluon, while giving 
higher integrated distribution, as shown in Fig.~\ref{dijet}. This effect
is caused by strong suppression of the GBW gluon at transverse 
momenta larger than the saturation scale, as we demonstrated
in Fig.~\ref{kmr-gbw}b.

In Fig.~\ref{gpbb-coll}a we show results for the total cross section,
$\sigma(\gamma p\to b\bar b X)$ obtained in the
$k_t$ factorization formalism and in the collinear
approximation, compared to the experimental data.
In both cases, the direct and resolved photon
contributions are added. The resolved photon contribution,
coming from the partonic process $q\bar q \to Q\bar Q
$is only evaluated in the collinear limit and demonstrated to
be negligibly small. In Fig.~\ref{gpbb-coll}b, the cross sections are
decomposed into the direct and resolved photon components.

\begin{figure}[ht]
\begin{center}
\epsfig{width= 0.75\columnwidth,file=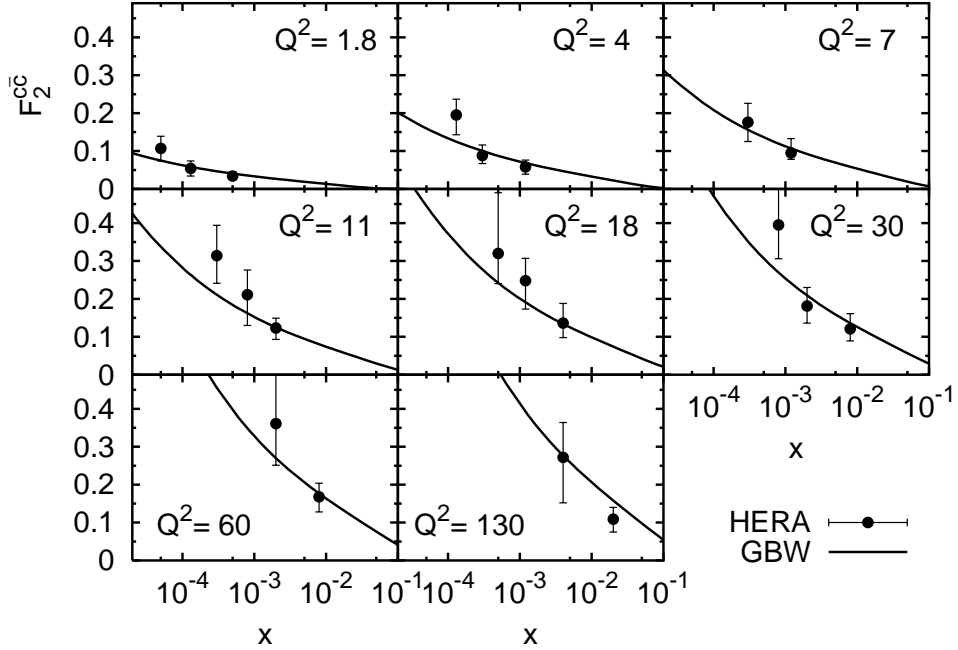,clip=}
\caption{The structure function $F_2^{c\bar{c}}$ as a function of $x$ at
different values of $Q^2$ between 1.8~GeV$^2$ and 130~GeV$^2$. The continuous
lines represent the results from the saturation model, with the parameters from
\cite{TKM}, compared with the experimental results from ZEUS \cite{exp_f2c}.
}\vspace*{-5mm}
\label{f2charm}
\end{center}
\end{figure}

We used the unintegrated gluon in the proton obtained via 
the KMR method (uGRV) and the
default values of the other parameters for the 
$k_t$~factorization calculation, as given in Sec.~\ref{sec4b}. The
collinear limit results are obtained using the corresponding
integrated gluon distributions (GRV), transverse momenta are set
to zero in the definition of energy scale for $\alpha_s$,
that is we take $\alpha_s(M^2)$. Other parameters take their
default values. Therefore, we gain insight into the
actual difference between the studied approximations, not
caused by discrepancy of input parameters. Thus, it is clear
from Fig.~\ref{gpbb-coll}b, that a large enhancement (by a factor of
three) of the resolved photon cross section occurs due to
non-zero transverse momentum effects, whereas the direct
photon contribution gains only  modest 10--20\% in the
magnitude. The total cross section is larger by some
20--30\% in the $k_t$ factorization approach.

It has been checked that charm production at HERA
is well described within the $k_t$ factorization approach\cite{charm}.
For completeness, we show in Fig.~\ref{f2charm} our results for the
charm structure function $F_2 ^{\mathrm charm}(x, Q^2)$,
based on the standard saturation model, compared to experimental
data. The quark mass $M_c = 1.3$~GeV.
The agreement of the theory and the data is very good,
in contrast to the discussed bottom production case.
The results we have obtained for bottom production in the single 
resolved case are compatible with previous studies of heavy quark
production using the $k_t$~factorization approach \cite{LSZ,Cristiano}.
These studies found agreement with the first erroneous data point from
HERA \cite{HERABB}, which gave a cross section for bottom production
lower by a factor 2 than found later \cite{HERABB}. With the
inclusion of the resolved photon in the $k_t$~factorization
framework, we restore a similar level of agreement with the experimental data.

\subsection{Results for $\gamma\gamma$ interactions}
\label{sec4c}

\begin{figure}[ht]
\begin{center}
\begin{tabular}{cc}
\raisebox{40ex}[0cm][0cm]{a)} & 
\epsfig{width= 0.5\columnwidth,file=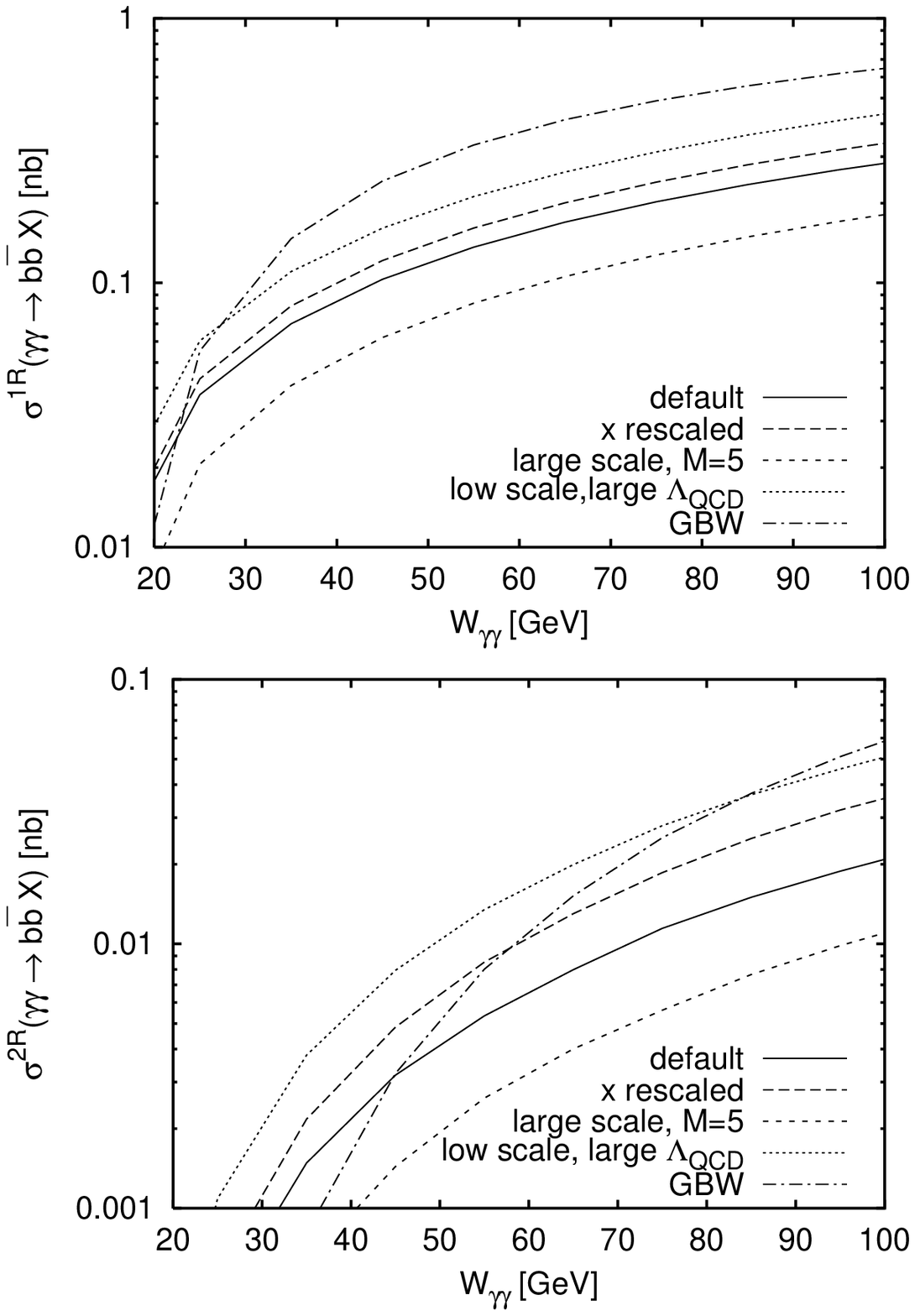} \\
\raisebox{10ex}[0cm][0cm]{b)} & 
\end{tabular} 
\caption{Cross sections of \bbbar production in $\gamma \gamma$ interactions
from the $k_t$~factorization approach, showing a) the single resolved
contribution and b) the double resolved contribution. Details of the
curves are presented in the text.
}
\label{ggbb}
\end{center}
\end{figure}

The default set of parameters for bottom production in
$\gamma\gamma$ collisions is the same as it was in the case
of the $\gamma p$ process. The unintegrated gluon in the
photon is unfolded from the GRS parameterization, using the
KMR method (see Sec.~\ref{sec2a}). We investigate the sensitivity of the
cross section to variations of the parameters. The results
are shown as a function of $\gamma\gamma$ energy $W$ in
Fig.~\ref{ggbb}. The default results (continuous line) are compared
with the results incorporating the kinematical rescaling
of~$x$ (see Sec.~\ref{sec4a}). Besides that, the {\em low} scale in
the running formula of $\alpha_s$ and the large $\Lambda_{\mathrm QCD}
$are assumed and the case of $M_b = 5.0$~GeV and the {\em large}
scale of $\alpha_s$ is shown. Furthermore, the unintegrated
gluon in the photon from the saturation model is used. We
stress that within the presented $W$~range, the results are
driven by the gluon at relatively large~$x \sim 0.1$, where
the saturation model is less reliable and phenomenological
threshold factors need to be imposed.

\begin{figure}[ht]
\begin{center}
\begin{tabular}{cc}
\multicolumn{2}{c}{\epsfig{width= 0.95\columnwidth,file=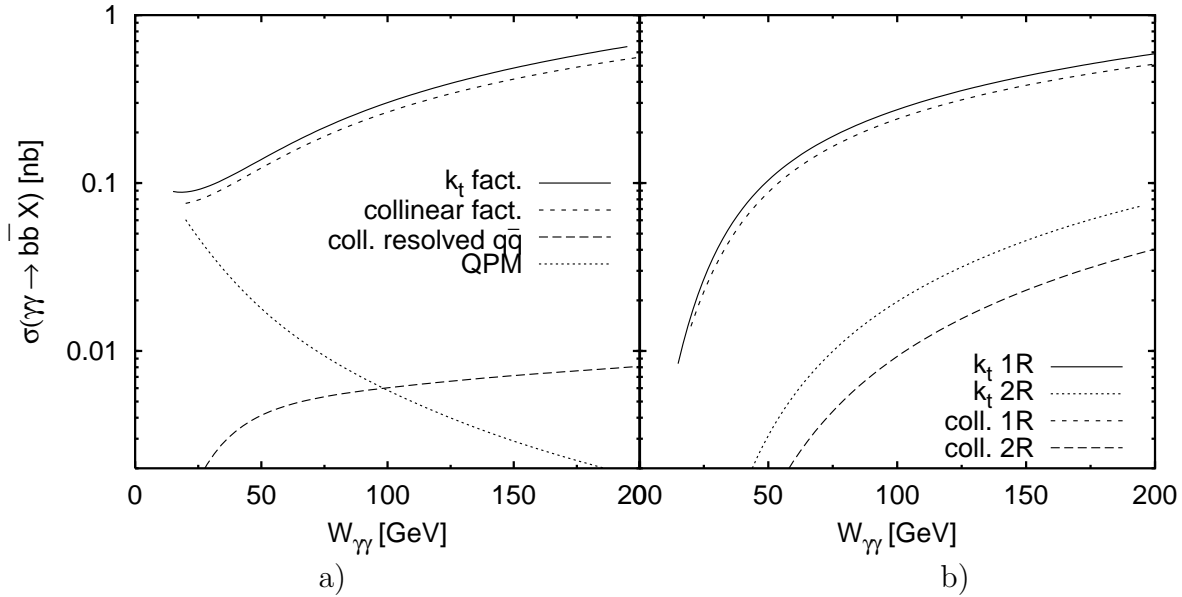}}\\
 \hspace{10em} a) & \hspace{10em}  b)
\end{tabular} 
\caption{Comparison between the collinear factorization and the 
$k_t$~factorization approaches to \bbbar production in $\gamma\gamma
$interactions.
a) The total \bbbar cross section is represented with the solid line
in the $k_t$~factorization and with the short dashed line in the
collinear factorization. The long dashed line gives the specific contribution 
from the $q\bar q$ annihilation in the double resolved case, corresponding
to the collinear approach. The dotted line shows the contribution from
the quark box (see Fig.~\ref{diagrams}a).
b) Comparison between the single resolved components in the
$k_t$ (solid line) and collinear (short dashed) factorizations,
and between the double resolved components in the $k_t$ (dotted)
and collinear (long dashed) factorizations.
}
\label{ggbb-coll}
\end{center}
\end{figure}

Furthermore in Fig.~\ref{ggbb-coll}, we illustrate the enhancement of the
cross sections due to non-zero transverse momentum of the gluon. This
figure is constructed in strict analogy to Fig.~\ref{gpbb-coll} discussed in
Sec.~\ref{sec4b}. The conclusions from these results are
also very similar to those obtained in the $\gamma p$ case.
Let us only mention, that the integrated and unintegrated
GRS parameterization of the gluon in the photon was used.
Note, that the QPM contribution (see the diagram in
Fig.~\ref{diagrams}a) shown in
Fig.~\ref{ggbb-coll}a is included in the total cross sections
$\sigma(\gamma\gamma \to b\bar b X)$.  
In order to obtain the total cross section 
$\sigma(e^+e^- \to e^+ e^- b\bar b X)$, the $\gamma\gamma$ cross 
section needs to be weighted with the photon flux in the electrons.
Thus, we obtained the value of 
$\sigma(e^+e^- \to e^+ e^- b\bar b X)=1.9$ pb for the default case
(based on KMR gluon density), while the use of the GBW gluons
gives $\sigma(e^+e^- \to e^+ e^- b\bar b X)=2.7$ pb. The latter value
is lower than what we previously found using a generalization of
the saturation model \cite{TKM}, with the difference coming from
the threshold factor imposed at the level of the
unintegrated gluon ${\cal F}_g (x, \tdm k^2)
$instead of the total cross section.
The behavior of the cross section in the vicinity of the kinematic
threshold is crucial for interpretation of the LEP measurements
of bottom production. An interesting discussion of this problem
may be found in \cite{Szczurek}.

The general picture, which emerges, may be summarized with
the following. The theoretical uncertainty of the QCD
results for bottom production is rather large --- of the
order of $50\%$ for the single resolved case and, even
larger for both photons resolved. The double resolved photon
contribution is only a small correction (a few percent) to
the single resolved photon over the studied energy range.
The QCD models give results 2 times smaller than the
GBW-model. The total cross section following from the
$k_t$~factorization scheme are not significantly larger than the
ones obtained in the collinear limit. Therefore, the
enhancement due to the use of unintegrated gluon is not
sufficient to solve the $b$-excess puzzle in $\gamma\gamma$ collisions.

\begin{figure}[ht]
\begin{center}
\begin{tabular}{cc}
\raisebox{40ex}[0cm][0cm]{a)} & 
\epsfig{width= 0.5\columnwidth,file=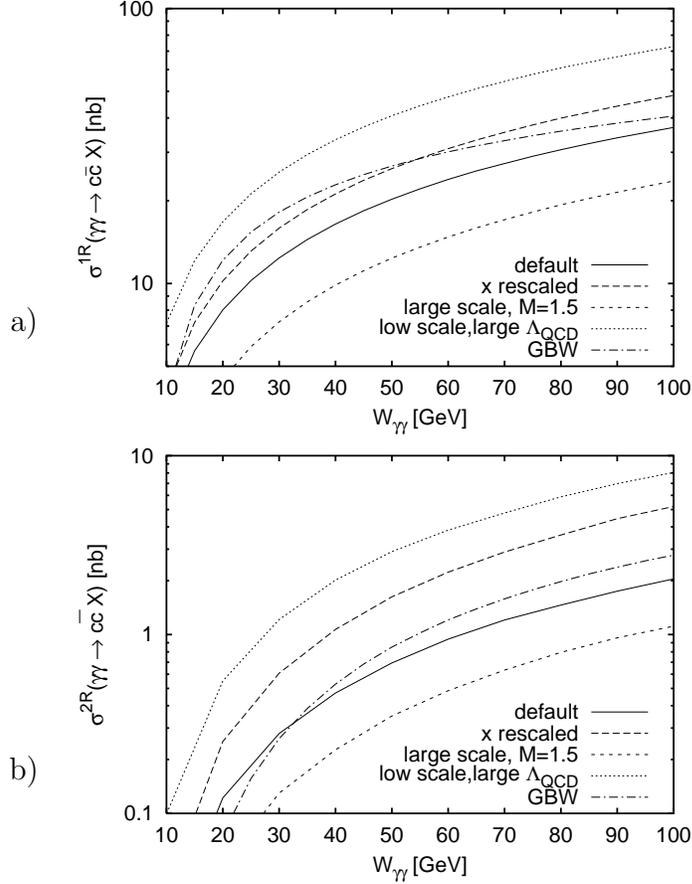} \\
\raisebox{10ex}[0cm][0cm]{b)} & 
\end{tabular} 
\caption{Cross sections of \ccbar production in $\gamma \gamma$ interactions
from the $k_t$~factorization approach, showing a) the single resolved
contribution and b) the double resolved contribution. Details of the
curves are presented in the text.
}
\label{ggcc}
\end{center}
\end{figure}

\begin{figure}[ht]
\begin{center}
\epsfig{width= 0.5\columnwidth,file=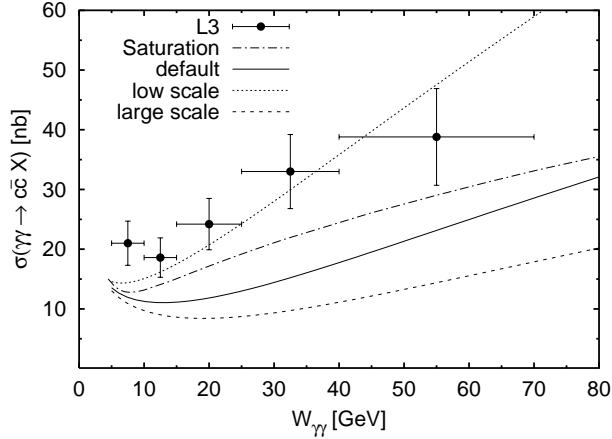}
\caption{Cross section for the inclusive charm production in $\gamma\gamma
$collisions. The data from the L3 experiment \cite{LEPCC} are compared with the
outcome of the $k_t$~factorization. The solid line represents our default 
choice for scales and parameterizations of the gluon in the photon, summing
all contributions. The dashed line and the dotted line give the uncertainty
due to the choice of scale. The dashed-dotted line represents our
previous calculation \cite{TKM} using the generalized saturation model.
}
\label{ggcc-coll}
\end{center}
\end{figure}

In an analogous scheme as for the bottom, we present in
Fig.~\ref{ggcc} the results for $\sigma(\gamma\gamma \to c\bar c
X)$. Because of the lower quark mass, the uncertainties
related to the energy scale choice and the definition of
gluonic~$x$ are larger than for the bottom production. The
double resolved photon contribution is again  small, few
percent correction to the dominant single resolved photon
contribution. The data points are fitted well within the
uncertainty band of theoretical estimates, as seen in
Fig.~\ref{ggcc-coll}. The extended GBW model 
as introduced in \cite{TKM} and
QCD give similar results for charm within the shown energy
range, whereas in the bottom case the GBW gluon
gave a larger cross section than QCD. 
The main reason of this difference is that, within the considered 
range of energies, the gluon is probed at lower~$x$ for charm, 
as compared to the bottom case.

\section{Discussion}
\label{sec5}

The main goal of this study is to investigate whether 
the puzzle of bottom production excess in $\gamma p$ and 
$\gamma \gamma$ collisions may be naturally explained in the
$k_t$~factorization framework. Thus, we have estimated the 
cross sections for direct and resolved photon(s) and 
compared the results to their counterparts in 
standard collinear approximations.

We, indeed, found some enhancement of the cross section 
for a direct photon scattering off both the proton and the resolved 
photon. This effect itself is too small, though, to get the theory results close
to the experimental data. Furthermore, we showed, that the 
cross sections driven by two-gluon fusion are substantially larger 
(by a factor of 2--3) when the unintegrated gluon is used. 
Despite this enhancement, these subprocesses contribute only
as a 20--30\% correction to the total cross section for
bottom production in the $\gamma p$ collisions, and less than
10\% for $\gamma\gamma$. 

The sensitivity of the results to various model uncertainties is 
found to be large. Thus, in the marginal case,
the theoretical results for $\gamma p \to b\bar b X$ in
the $k_t$~factorization framework, is enhanced and agrees with 
the H1 data point within errors.  
We do not interpret this as a strong indication of 
consistency between data and the theory, but rather as a 
consequence of the wide uncertainty band. 
Better understanding of higher order corrections is crucial
to determine whether the H1 data point contradicts
QCD results. 

The picture is much more clear for $\gamma\gamma \to b\bar b X$.
This process is dominated by the QED box diagram at low energies
and by the single resolved photon mechanism at larger energies. 
Irrespectively of the assumption made within our framework, the 
emerging results for $e^+e^- \to e^+ e^- b\bar b X$,
$\sigma^{th}= 1.9$~pb at the $e^+e^-$ collision energy 
$\sqrt{s}=200$~GeV,
 are more than three standard deviations below the experimental 
data,
$13.1 \pm 2.0 \, {\rm (stat)} \pm 2.4 \,{\rm (syst)}$~pb \cite{LEPHQ}.
Thus, in the standard QCD+QED approach, the $b$~production
in $e^+ e^-$ collisions is hard to explain.

Interestingly enough, the agreement between the theory 
and the experimental data is good for the charm case both
for $\gamma\gamma$ and $\gamma^* p$. 
The very different behavior of charm and bottom cross sections
is surprising and calls for an explanation.

\section{Conclusions}
\label{sec6}

In this study we have analyzed photoproduction of heavy flavors in 
$\gamma\gamma$ and $\gamma p$ collisions using the $k_t$~factorization
approach. First, we obtained parameterizations of unintegrated
gluon in the photon using the KMR method. We compared 
features of the unintegrated gluon in the photon and in the proton.
Furthermore, we proposed a parameterization of the gluon in
the photon based on a generalization of the saturation model.

The parameterizations, combined with off-shell matrix elements, 
were used to estimate cross sections for charm and bottom production,
including contributions from resolved photon. The impact of
non-zero transverse momentum of gluons was studied. Some enhancement
of the cross section was found in the $k_t$~factorization approach.
In particular, we demonstrated the importance of the resolved 
photon contribution to bottom production in $\gamma p$ collisions.
Sensitivity of the theoretical estimates to the details of
the model was investigated. 
The conclusion is, that the use of $k_t$~factorization
approach brings the theoretical results for bottom 
production closer to the data, but large discrepancies 
remain. For $\gamma p$ collisions, a major inconsistency 
cannot be claimed, because of large experimental errors 
and theoretical uncertainties. For the $\gamma\gamma$ case,
the $b$-production excess is statistically significant,
despite the uncertainties. The $k_t$ approach, based on QCD,
does not agree with the data from LEP at the three sigma level. 
On the contrary, the charm production is well understood 
within QCD. 

Thus, an interesting question arises, why the 
$b$-quark production excess is found in various
processes. A potential explanation may be, perhaps,
provided by higher order corrections. Still, 
it is not clear why similar, or even more important, 
corrections would not affect charm production in a 
similar way. It is also interesting to ask about
uncertainties of parton densities in the photon
since the experimental constraints from measurements 
of the photon structure and a jet photoproduction are
not very stringent. 
Definitely, an attempt should be made to perform 
a new global fit of parton densities in the photon,
including the bottom production data.
Another interesting explanation of the $b$-excess at the 
Tevatron was suggested in \cite{Cacciari} where 
the non-perturbative fragmentation function of the 
$b$~quark into $B$~mesons was updated by fitting to precise 
LEP data. It was found, that the discrepancy between 
standard theoretical calculations and the Tevatron data is 
significantly reduced when the improved model of fragmentation 
is used.
Finally, there is also a possibility, 
that the answer to the question leads beyond QCD\cite{sbottom}. 
Therefore, it is important to further constrain 
theoretical uncertainties of bottom production 
rate estimates in QCD.

\section*{Acknowledgements}
We are very grateful to Gunnar Ingelman, Jan Kwieci\'{n}ski
and Johan Rathsman  for valuable
discussions during the course of this research. 
We are grateful
to Matteo Cacciari for interesting correspondence about
the role of fragmentation function for $b$~physics.
This research was supported by the Swedish Research Council
and by the Polish Committee for Scientific Research (KBN) 
grant no. 5P03B 14420.

\newpage

\appendix
\section{Virtual gluon-photon fusion}
\label{apB}
The partonic cross section for an off-shell gluon
with the transverse momentum $\tdm k
$depend on the density matrix for gluon
polarizations  $\varepsilon^{\mu} (\tdm k)$.
In this study, we assume that the gluons have
the BFKL-like polarization tensor, that is
\be
\label{ap:polarization}
\sum_{\lambda} \epsilon^{\mu} _{\lambda}
{\epsilon^{\nu} _{\lambda}}^* =
{\tdm k^{\mu}\tdm k^{\nu} \over \tdm k^2}.
\ee
The result of the integration over the phase for an off-shell
matrix element describing a gluon-photon fusion 
$\gamma(p) g^*(k) \to Q\bar Q$ is well
known. We quote the result from \cite{CCH}
\[
\hat\sigma^{kf} _{\gamma g}(\tdm k^2,M^2,\nu) =
{\pi \alpha_{em} e_Q^2 \alpha_s(\bar \mu^2) \over 2M^2}
\Theta(\nu - 4M^2 - \tdm k^2) \rho \beta
\left\{  
[(1+\rho-\frac{1}{2}\rho^2)\, L(\beta) - 1 - \rho ] 
\right.
\]
\be
\left.
+ [8+\rho-(2+3\rho) L(\beta)] \frac{\tdm k^2}{\nu} +
 [-8+2 L(\beta)] \left( \frac{\tdm k^2}{\nu} \right)^2
\right\}
\label{gamgluqq}
\ee
with
\be
\nu = 2\, p \cdot k, \qquad \rho = {4M^2 \over \nu},
\ee
\be
\beta = \sqrt{1 - \rho \left( 1 - {\tdm k^2 \over \nu}
\right) ^{-1}}
\ee
and
\be
L(\beta) = {1\over \beta} \log {1+\beta \over 1-\beta}.
\ee
The heavy quark charge is denoted by $e_Q$.

\section{Cross sections for virtual gluons}
\label{apA}

In this appendix, we follow conventions of \cite{BE}, with
some minor modifications.
The kinematics of the gluon fusion process
$g^*(k_1)+ g^*(k_2) \longrightarrow Q(p_4) + \bar{Q}(p_3)$, with respect to
the four-vectors $p_1$ and $p_2$ of the incoming photons/hadrons, are given by
\ba
\label{ap:kinematics}
k_1 & =  & x_1 p_1 + \tdm k_1 , \nonumber \\
k_2 & =  & x_2 p_2 - \tdm k_2 , \nonumber \\
p_3 & =  & (1-z_1) x_1 p_1 + z_2 x_2 p_2 + \tdm k_1 - \tdm \Delta , \nonumber \\
p_4 & =  & z_1 x_1 p_1 + (1-z_2) x_2 p_2 - \tdm k_2 + \tdm \Delta .
\ea

The two-body phase space $d\Phi^{(2)}$ of the $Q\bar{Q}$ pair can be written in
the following way
\be
\label{ap:phasespace}
d\Phi^{(2)} = \frac{1}{8\pi^2}\frac{dz_1}{z_1(1-z_1)} \,
d^2\tilde{\tdm \Delta} \,
\delta\left(\nu-\frac{\tilde{\tdm \Delta}^2+M^2}{z_1(1-z_1)}-\tdm q^2\right) ,
\ee
where
\ba
\label{ap:definitions}
\tdm q & = & \tdm k_1 - \tdm k_2 , \nonumber \\
\nu & = & x_1 x_2 (2p_1 \cdot p_2) = \hat s + \tdm q^2 , \nonumber \\
\tilde{\tdm \Delta} & = & \tdm \Delta - \tdm k_1 z_1- \tdm k_2 (1-z_1) .
\ea

Furthermore, the following relations hold
\ba
\label{ap:relations}
\hat s &=& \frac{\tilde{\tdm \Delta}^2+M^2}{z_1(1-z_1)} , \nonumber \\
M^2-\hat t &=& z_1(\hat s+{\tdm k_1}^2)+(1-z_1){\tdm k_2}^2
+2\tilde{\tdm \Delta}\cdot{\tdm k_2} , \nonumber \\
M^2-\hat u &=& (1-z_1)(\hat s+{\tdm k_1}^2)+z_1{\tdm k_2}^2
-2\tilde{\tdm \Delta}\cdot{\tdm k_2} , \nonumber \\
z_2 &=& \frac{[(1-z_1)\tdm q-\tilde{\tdm \Delta}]^2+M^2}{(1-z_1)\nu} ,
\nonumber \\
1-z_2 &=& \frac{[z_1 \tdm q+\tilde{\tdm \Delta}]^2+M^2}{z_1 \nu} .
\ea

In this notation, the cross section 
$\hat \sigma^{kf} _{gg}(\tdm k_1, \tdm k_2, M^2, \nu)
$for production of heavy quarks in two-gluon collisions
$g^* g^* \to Q\bar Q$ equals (from \cite{BE})
\be
\label{ap:cross-section}
\hat \sigma^{kf} _{gg}
(\tdm k_1, \tdm k_2, M^2, \nu) =
\frac{4 \pi^2}{N_c^2-1} \nu \alpha_s^2 (\bar \mu^2) 
\int d\Phi^{(2)} D(\tdm k_1, \tdm k_2, \nu, \tdm \Delta, z_1, M^2),
\label{glugluqq}
\ee
and the matrix element $D(\tdm k_1, \tdm k_2,\nu,\tdm \Delta, z_1, M^2)$ 
is given by
\ba
\label{ap:matrixelement}
D &=& \frac{1}{N_c}\left[\frac{-1}{(M^2-\hat u)(M^2- \hat t)}+
\frac{(B+C)^2}{{\tdm k_1}^2 {\tdm k_2}^2}\right] + \nonumber \\
   &+& N_c \left[\frac{1}{\hat s}\left(\frac{1}{M^2-\hat t}-
\frac{1}{M^2-\hat u}\right) (1-z_1-z_2)- \frac{B^2+C^2}{{\tdm k_1}^2
{\tdm k_2}^2}\right.+ \nonumber \\
&&\hphantom{N_c \Bigg[}\!\left.+ \frac{2(B-C)}{{\tdm k_1}^2
{\tdm k_2}^2 \hat s}\left((1-z_2) {\tdm k_1}^2 +(1-z_1) {\tdm k_2}^2
+{\tdm k_1} \cdot{\tdm k_2}\right)\right] + \nonumber \\
  &+&N_c \left[\frac{2}{\nu \hat s}-\frac{2}{{\tdm k_1}^2 {\tdm k_2}^2}
\frac{\left((1-z_2) {\tdm k_1}^2+(1-z_1){\tdm k_2}^2
+{\tdm k_1}\cdot{\tdm k_2}\right)^2}{\hat s^2}\right] ,
\ea
where the following notation has been used
\ba
\label{ap:notation}
B&=&\frac{1}{2}-\frac{(1-z_1)(1-z_2)\nu}{M^2-\hat u}+
\frac{\nu(1-z_1-z_2)}{2\hat s}+
\frac{\tdm \Delta \cdot({\tdm k_1}+{\tdm k_2})}{\hat s} , \nonumber \\
C&=&\frac{1}{2}-\frac{z_1z_2\nu}{M^2-\hat t}-\frac{\nu(1-z_1-z_2)}{2\hat s}-
\frac{\tdm \Delta \cdot({\tdm k_1}+{\tdm k_2})}{\hat s} .
\ea

\end{document}